\documentclass[twocolumn,secnumarabic,showpacs,amssymb,amsmath, nobibnotes, aps, prl, a4paper]{revtex4-1}
\usepackage{bm}
\usepackage{amsmath}
\usepackage{graphicx}  

\providecommand{\EE}[2]{\relax\ifmmode #1\times10^{#2} \else $#1\times10^{#2}$\fi}

\newcommand{\TD}{T_\mathrm{D}}
\newcommand{\kD}{k_\mathrm{c}}
\newcommand{\fD}{f_\mathrm{D}}

\newcommand{\AD}{A_\mathrm{D}}
\newcommand{\bAD}{\bar{A}_\mathrm{D}}
\newcommand{\TtwoB}{T^{\rm 2B}}
\newcommand{\wrho}{\omega_\rho}
\newcommand{\wz}{\omega_z}
\newcommand{\TR}{T_{\rm R}}
\renewcommand{\vr}{\vec{r}}

\begin{document}

\title{Observation of a space-time crystal in a superfluid quantum gas}

\author{J. Smits${}^1$}
\author{L. Liao${}^2$}
\author{H.T.C. Stoof${}^2$}
\author{P. van der Straten${}^1$}
\email[]{p.vanderstraten@uu.nl}

\affiliation{$^1$Debye Institute for Nanomaterials and Center for Extreme Matter and Emergent Phenomena, Utrecht University, PO Box 80.000, 3508 TA Utrecht,The Netherlands\\
$^2$Institute for Theoretical Physics and Center for Extreme Matter and Emergent Phenomena, Utrecht University, PO Box 80.000, 3508 TA Utrecht,The Netherlands}
\date{\today}%

\begin{abstract}
Time crystals are a phase of matter, for which the discrete time symmetry of the driving Hamiltonian is spontaneously broken.  The breaking of discrete time symmetry has been observed in several experiments in driven spin systems. Here, we show the observation of a space-time crystal using ultra-cold atoms, where the periodic structure in both space and time are directly visible in the experimental images. The underlying physics in our superfluid can be described {\em ab initio} and allows for a clear identification of the mechanism that causes the spontaneous symmetry breaking. Our results pave the way for the usage of space-time crystals for the discovery of novel nonequilibrium phases of matter. 
\end{abstract}

\maketitle

Frank Wilczek proposed the idea of time crystals in 2012~~\cite{wilczek_tc}, where in analogy to space crystals the continuous time symmetry  is broken spontaneously. Since that time there has been discussion on what should constitute a time crystal~\cite{bruno,nozieres} and how to create them. Watanabe \emph{et al.}~\cite{oshikawa_no_tc} showed that in principle the continuous time symmetry cannot be broken spontaneously into a discrete symmetry in the ground state. However, there have been proposals to realize instead a discrete time crystal by breaking of a discrete time translation symmetry~\cite{nayak_floq_tc,sacha_floq_tc,schoen_ph_sp_xtal,sondhi1,sondhi2}. Following a theoretical model by Yao \emph{et al.}~\cite{yao} several experiments~\cite{monroe_exp,lukin_exp,Rovny,sreejith} realized this particular symmetry breaking in driven spin systems. These experiments were limited to probing a very restricted number of particles~\cite{monroe_exp} or an ensemble of particles without any spatial resolution~\cite{lukin_exp,Rovny,sreejith}, preventing the direct observation of spatial ordering. 

In this Letter, we report the direct observation of a space-time crystal exhibiting not only periodic oscillations in time with double the period of the driving force, but also an oscillatory spatial structure, \emph{i.e.}, both a discrete time translation symmetry as well as the continuous spatial translation symmetry are broken. Due to the small dissipation in our superfluid gas we can study the space-time crystal over an extensive period of time showing  the collapse and revival of the oscillating long-lived spatially ordered state. 
Superfluid quantum gases are the ideal system to study discrete time-crystals. Due to the low viscosity and heat conduction, excitations in the system can be induced without the associated heating of the system. Periodic driving of the excitations in the system can easily be arranged due to the harmonic confinement of the atoms in the trap. Crucial in the driven spin systems~\cite{monroe_exp,lukin_exp,barrett_exp} has been the occurrence of strong disorder, where either many-body localization or some other mechanism is the cause for the small dissipation in the experiments. 

\begin{figure}
	\includegraphics[width=\linewidth]{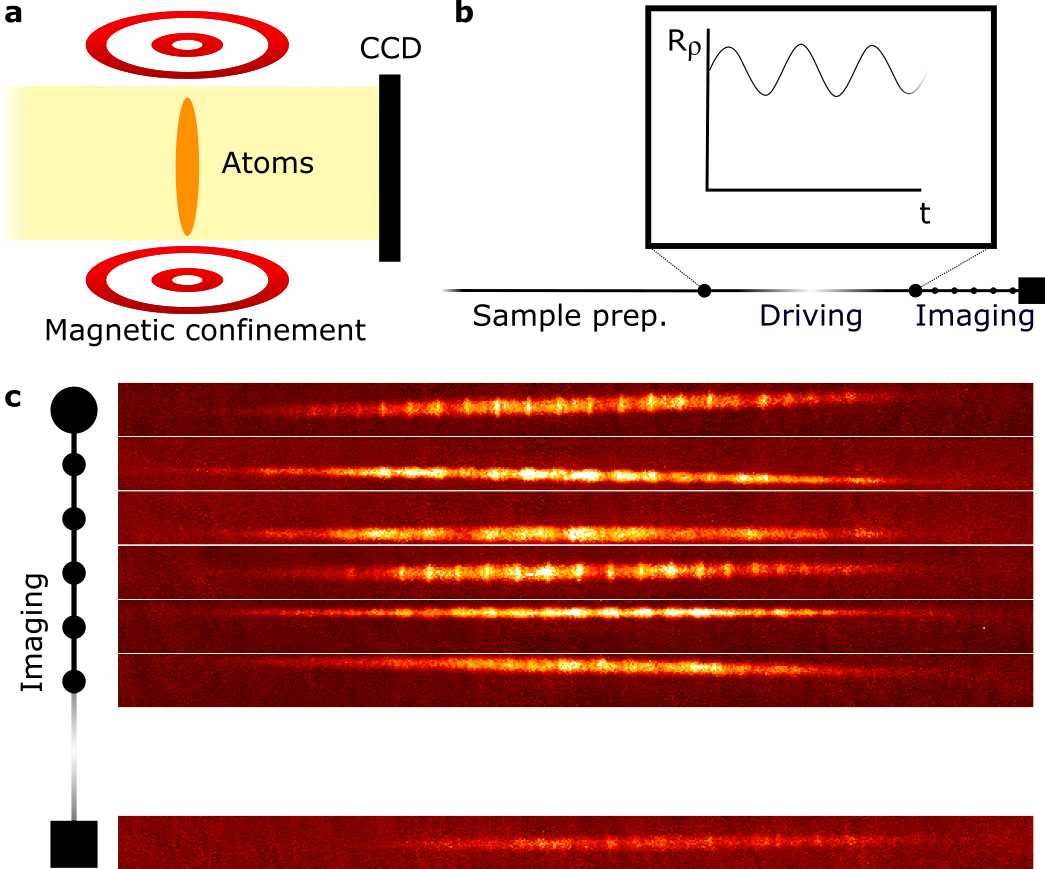}
	\caption{(color online). Schematic representation of the experimental setup, the timeline, and the imaging sequence. \textbf{(a)} Schematic view of imaging system and atomic cloud. \textbf{(b)} Experimental sequence and schematic representation of the driving. \textbf{(c)} Imaging sequence showing the first six and the last image of a selected run. In total 50 images are taken each run with $3.28\,\mathrm{ms}$ between images. }
	\label{figone}
\end{figure}

However, as shown by Else \emph{et al.}~\cite{else}, time-crystals can also exist in the prethermal regime, if the drive frequency is sufficiently large compared to the excitation frequency. Following these experiments there have been a large number of proposals~\cite{Huang,Sacha,lustig,WenWei,Mizuta,Russo} for the observation of time crystals using several different systems (see also the review~\cite{sacha-review}). In superfluid quantum gases disorder is absent.  Since superfluid quantum gases can be imaged using phase-contrast techniques, which allows the accumulation of several tens of images of the same superfluid cloud, the dynamics of the system can be studied over many cycles. Moreover, as the conditions of the space-time crystal are not very sensitive to the initial drive of the excitations, the  superfluid cloud can be studied over a prolonged period of time by combining multiple measurement series together extending the observation period to several seconds. Finally, the dynamics of the superfluid quantum gas in a radial symmetric trap can be simulated using time-splitting spectral methods~\cite{bao}, which allows us to compare our experimental findings with simulations to elucidate the mechanisms behind the space-time crystal formation.

The superfluid is produced in the trap in a cigar-shaped form, where the ratio between the trap frequencies causes the axial size to be about 40 times larger than the radial size. After sample preparation~cite{refmat}, the radial trap frequency is suddenly perturbed and this induces a radial breathing mode of the cloud with a frequency of $\fD$ = 104.691(16) Hz, which is only weakly damped and has a decay time of several seconds. This radial breathing mode with a period $\TD=1/\fD$ acts as the drive for the excitation of the cloud in the axial direction. After many radial oscillations a high-order excitation emerges in the axial direction, which has been observed previously and interpreted in that paper as ``Faraday waves''~\cite{engels_faraday}. By observing the spatio-temporal long-range order, we show that an interpretation as a space-time crystal is more appropriate using the modern language of nonequilibrium phase transitions. Figure~\ref{figone} shows several images of the pattern displaying the large variety in radial size and axial excitation. This axial pattern is only observed, if the radial breathing mode is strongly excited and the perturbation of the cloud is in the non-linear regime. 

\begin{figure}
	\includegraphics[width=\linewidth]{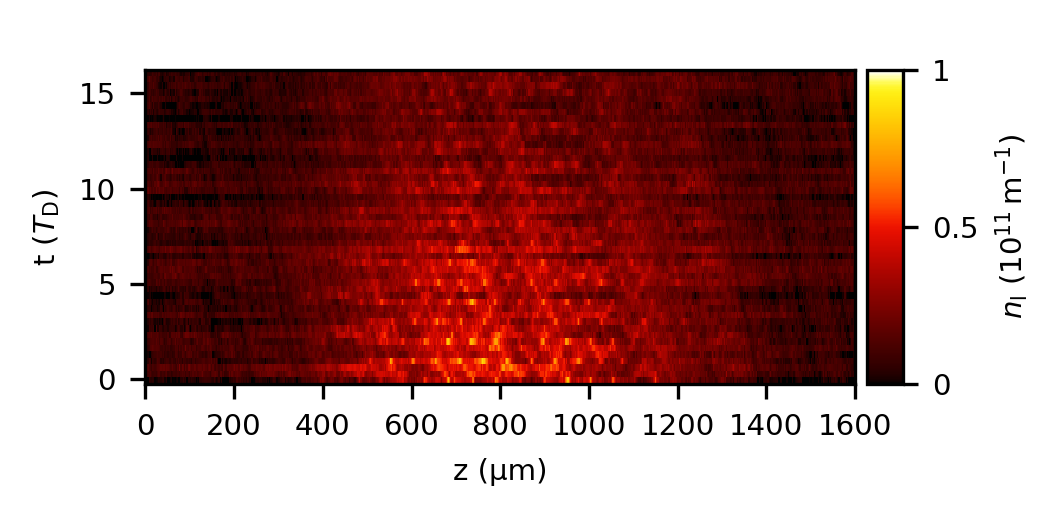}
	\caption{(color online). Line density $n_\ell$ as a function of time and position starting $500\,\mathrm{ms}$ after the onset of the drive. Time slices taken from a single experimental run. Both in space and time a recurring pattern is observed. The (temporal) period of the pattern corresponds to twice  the breathing period $\TD$. The diagonal streaks in the image are caused by correction for the uncoupled axial center-of-mass motion and darker areas in the imaging. Decrease of the signal is attributed to slight particle loss ($3\%$ per shot) due to interactions of imaging light with the atoms. }
	\label{figtwo}
\end{figure}

To study the axial pattern, the density profile is integrated over the radial direction and the result is shown in Fig.~\ref{figtwo} as a function of time. A lattice of maxima in the density is observed in both the temporal and spatial direction; a clear signature of a space-time crystal. The wavenumber of the pattern increases slightly towards the edges of the superfluid, which is attributed to the finite extent of the cloud. The period of the pattern is determined to be almost $2 \, \TD$ over the entire detection period and this sub-harmonic response to the drive is a requirement for the symmetry breaking implied by a discrete time crystal.

\begin{figure*}
	\includegraphics[width=\linewidth]{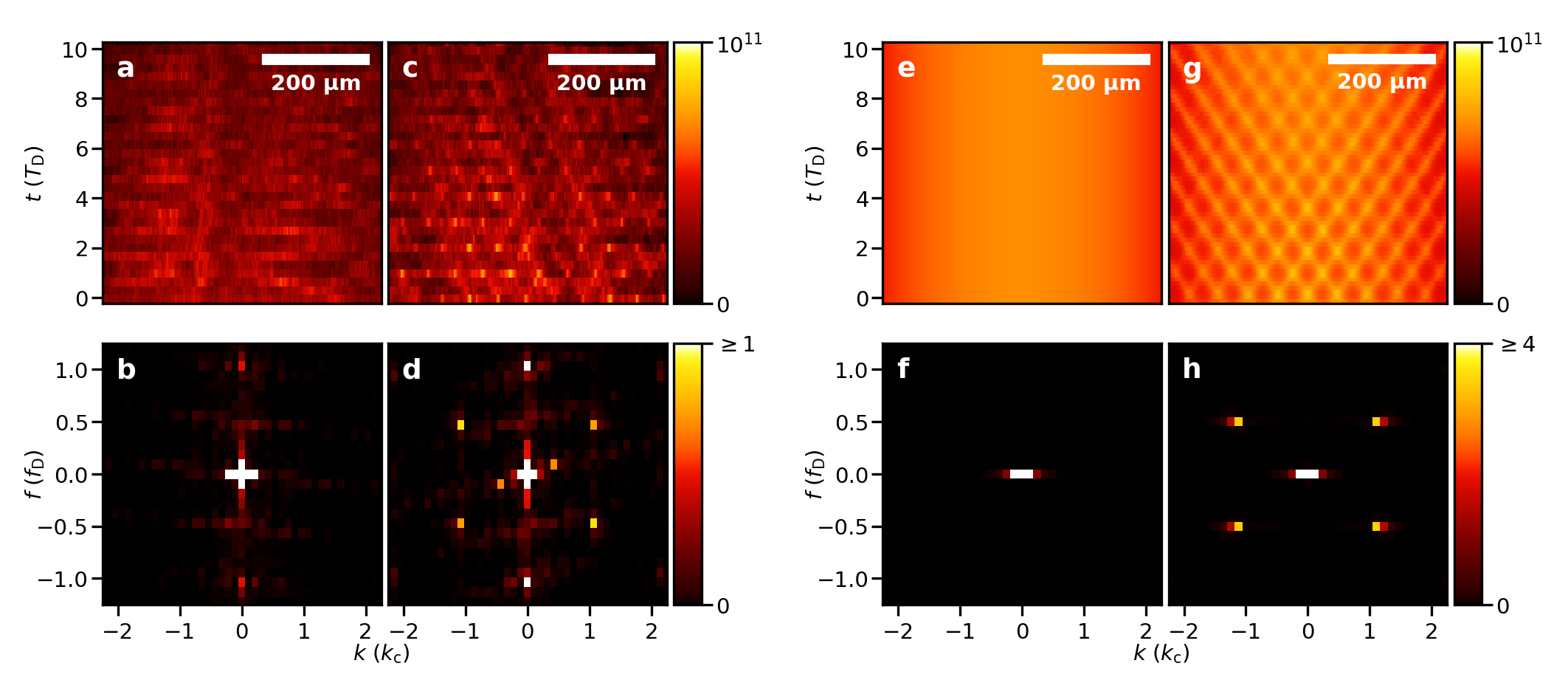}
	\caption{(color online). Fourier analysis and comparison of experiment with simulation. \textbf{(a)} Line density at the center of the cloud before the onset of the space-time crystalline phase, directly after the excitation. \textbf{(b)} Fourier transform of the data in (a). Peaks at $f/\fD=\pm 1$ are associated with a weakly excited scissors mode. The signal around the origin is associated with the equilibrium profile of the condensate. \textbf{(c)} Take-out of Fig.~\ref{figtwo}. Line density at the center of the cloud after the transition to the space-time crystalline phase, after a driving time of $500\,\mathrm{ms}$. A lattice has formed. \textbf{(d)} Fourier transform of the data in (c), with the appearance of four additional peaks due to the space-time crystal at $(k/\kD,f/\fD) = (\pm 1,\pm 0.5)$, where $\kD$ is the center wavelength~\cite{refmat}. \textbf{(e)} Simulated line density for a modulation depth of 0.02, after a wait time of $25\,\TD$. \textbf{(f)} Fourier transform of the data in (e). Notice that in the simulation only the equilibrium profile is visible. \textbf{(g)} Simulated line density for a modulation depth of 0.2 after a wait time of $25\,\TD$. A pattern similar  to the experimental data of (c) is observed. \textbf{(h)} Fourier transform of (g). Note the appearance of the four additional peaks at $(k/\kD,f/\fD) = (\pm 1,\pm 0.5)$ attributed to the space-time crystal. Line density in (a), (c), (e) and (g) is in units of $10^{11}$ atoms/m. Fourier images in (b), (d), (f), and (h) are truncated and normalized to 1 for the experimental data. }
	\label{figthr}
\end{figure*}

In Fig.~\ref{figthr}a,c the central part of the axial profile of Fig.~\ref{figtwo} is shown just after the start of the drive (Fig.~\ref{figthr}a) and after the axial excitation pattern emerged (Fig.~\ref{figthr}c). Figure~\ref{figthr}c shows that the space-time crystal has a centered cubic lattice structure with a period $2\,\TD$ in time. To determine the long-range temporal and spatial order, these patterns are Fourier transformed and shown in Fig.~\ref{figthr}b,d, respectively. The Fourier signal for the axial excitation pattern in Fig.~\ref{figthr}d contains four Fourier peaks at $(k/\kD,f/\fD)=(\pm 1,\pm 1/2)$, where the temporal frequency is half the driving frequency $\fD = 1/\TD$. This again shows that we are dealing with a discrete time crystal. The spatial periodicity  $2\pi/\kD$ is 57.3 $\mu$m as determined from the axial mode that we excite~\cite{refmat}. The appearance of the narrow peaks in the (momentum-frequency) Fourier plane is a clear indication of the simultaneous spatial and temporal long-range order in our system and manifestly indicates that we can truly speak of a space-time crystal. The Fourier signals in Fig.~\ref{figthr}b,d also contain two peaks in the temporal signal for non-zero frequencies at $f\simeq\pm\fD$ indicating the excitation of a weakly excited scissor mode. Such a mode can easily be induced due to small imperfections in the fabrication of the magnetic trap. 

In order to further check the validity of our experimental findings, we have numerically simulated the evolution of a Bose-Einstein condensation using a time-splitting spectral method under the same conditions regarding the number of atoms, the trap frequencies, and the drive assuming a radial-symmetric trap ~\cite{refmat}. The results are shown in Fig.~\ref{figthr}e-h and show excellent agreement with the experimental results apart from the weak scissor mode, which is absent in the simulations. This agreement shows that the physics of the space-time crystal for our experimental conditions is fully encapsulated in the Gross-Pitaevskii equation.

\begin{figure}
	\includegraphics[width=\linewidth]{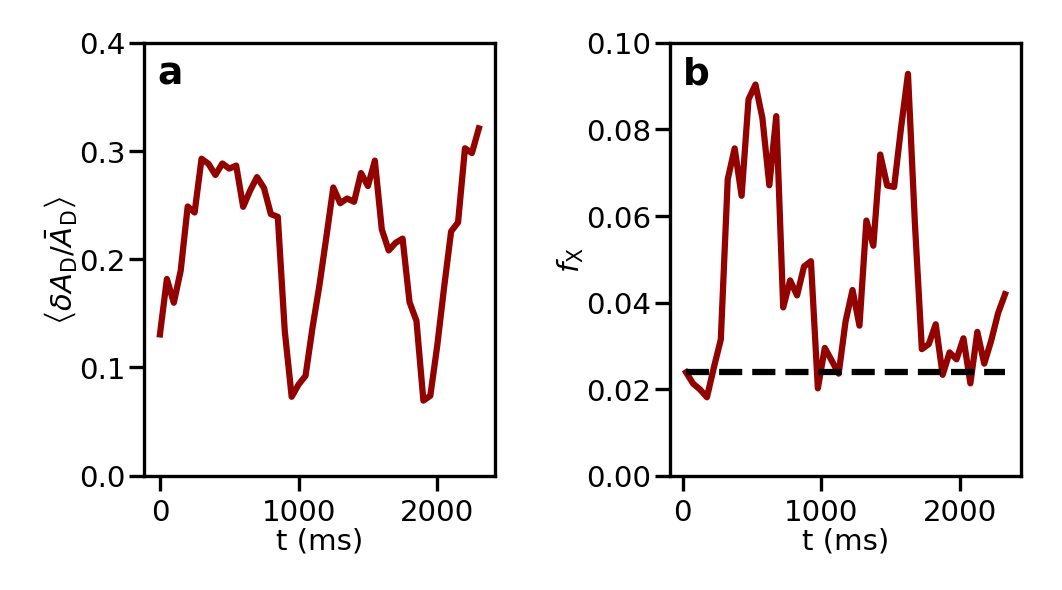}
	\caption{(color online). Long-term behavior of the amplitudes of the drive and crystal fraction. \textbf{(a)} Relative amplitude of the radial breathing mode derived by fitting a two-dimensional profile to data. \textbf{(b)}  Crystal fraction determined from each measurement run. The dashed line indicates background signal from shot-to-shot noise. Notice that the crystalline phase appears a certain time after the driving mode revives~\cite{refmat}. }
	\label{figfou}
\end{figure}

To demonstrate longevity of the space-time crystal, we compare the amplitude of the driving mode to the crystal fraction. The amplitude of the drive and emergence of the crystalline phase are shown in Fig.~\ref{figfou}. Over a full experimental run of $2.6\,\mathrm{s}$, the pattern is seen to appear and disappear two times. Appearances of the space-time crystal occur at times $t$ = 350 and 1350 ms, while the disappearance of the crystalline phase coincides with the decrease of the driving mode amplitude to near zero. The space-time crystal lasts, in each individual appearance, for over $500\,\mathrm{ms}$ or $50\,\TD$. The decrease of the driving mode is caused by the coupling to the scissor mode. The periodicity in the occurrence of the space-time crystal coincides approximately with the period that we extract from our simplified model describing the coupling between scissor and breathing mode~\cite{refmat}. The scissor mode has a period of about $\TD/2$ and is not linearly coupled to the axial excitation pattern due to parity conservation. 

Theoretically, we treat the space-time crystal variationally as a multimode system with the mode functions $P_{4j+2}(\tilde{z}) - P_{4j}(\tilde{z})$ with $\tilde{z}=z/R_z(t)$ in terms of Legendre polynomials, and frequencies $f_j$ excited by the drive due to the time-dependence of the Thomas-Fermi radii $R_x(t)$, $R_y(t)$ and $R_z(t)$, for which $R_i(t) = R_i(t+\TD)$ and $i$ = $x$, $y$, and $z$. After substituting this {\it ansatz} in the action for the Gross-Pitaevskii equation and neglecting nonlinear mode coupling, we ultimately obtain the Hamiltonian
\begin{equation}
\hat{H} = \sum_j \left[  2 \pi \hbar f_j a_j{}^\dag a_j + g_j(t) (a_j{}^\dag a_j{}^\dag+a_j a_j + 2 a_j{}^\dag a_j) \right],
\label{eq:model}
\end{equation}
where $a_j{}^{(\dag)}$ are the annihilation (creation) operators for quanta in the mode $j$ and $g(t)$ is the coupling with the periodicity of the drive. 
By moving to the rotating frame and applying the rotating-wave approximation to eliminate the time-dependence of the drive $g(t)$ we find the effective Hamiltonian
\begin{equation}
	\hat{H}_\mathrm{eff} = \sum_j  \left[ 2 \pi \hbar(f_j- \fD/2) a_j{}^\dag a_j + g_{j,0} (a_j{}^\dag a_j{}^\dag+a_j a_j) \right],
	\label{eq:ad_elim}
\end{equation}
where $g_{j,0}$ is proportional to the amplitude of the drive. Note that this yields a Hamiltonian, which is time independent in the rotating frame, and that represents the appropriate Hamiltonian for prethermalization of the system. 

\begin{figure}
	\includegraphics[width=\linewidth]{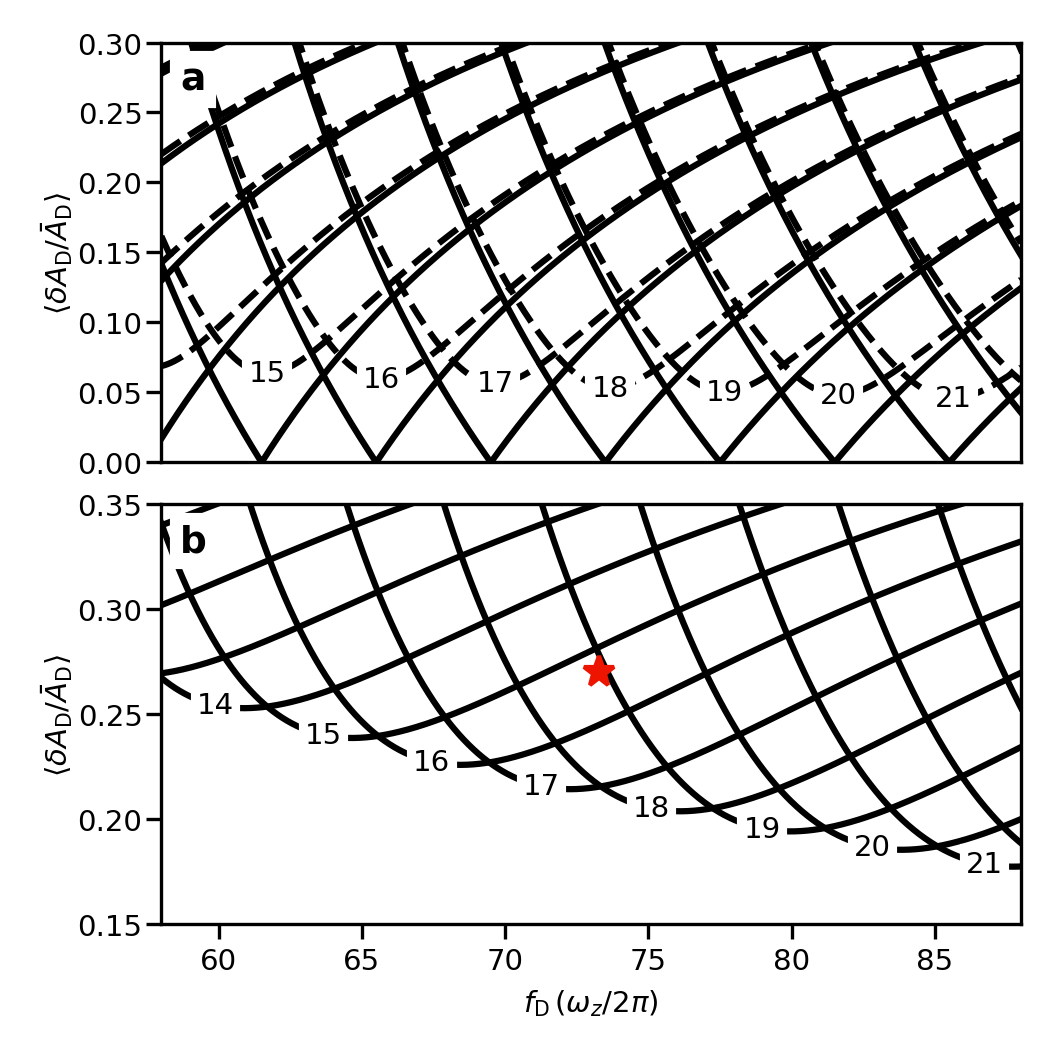}
	\caption{(color online). Minimum required driving amplitude $\langle \delta \AD/\bAD \rangle$ as a function of the driving frequency $\fD$ in linear response analysis~\cite{tobepublished}.  \textbf{(a)} Without damping (solid line) a mode $j$ can be driven with arbitrary small amplitude, if the driving frequency coincides with twice the mode frequency $f_j$, whereas for small damping (dashed line) there is for any drive frequency a threshold, below which the mode is not excited.     \textbf{(b)} Under our experimental conditions (indicated by the red star), the damping is larger and the threshold for exciting modes becomes larger. In the case of linear response, the modes $j$ = 16, 17, and 18 can be excited and depending on the competition between these modes, one of them dominates the pattern~\cite{tobepublished}. }
	\label{figfiv}
\end{figure}

The mode that is observed depends on the driving frequency $\fD$ and the driving amplitude $\langle \delta \AD/\bAD \rangle$~\cite{tobepublished}. In Fig.~\ref{figfiv} the minimum required amplitude is shown as a function of the driving frequency. In absence of damping, as shown in Fig.~\ref{figfiv}a, a mode $j$ can be driven with an arbitrary small amplitude, if the resonance condition $2 f_j =  \fD$ is fulfilled. In the case of damping, the threshold for exciting the pattern becomes finite.  Applying the analysis of Ref.~\cite{tobepublished} to our experimental conditions (see Fig.~\ref{figfiv}b) shows that the driving amplitude used in our experiment is sufficient to excite several modes $j$ and the competition between these modes causes one of the modes to grow exponentially and thus dominating the observed pattern.

The Hamiltonian of Eq.~(\ref{eq:ad_elim}) explicitly breaks the $U(1)$ symmetry $a\rightarrow a e^{i\vartheta}$. This implies that in the laboratory frame $\langle a_j a_j \rangle \propto e^{-2 \pi i \fD t}$ is always non-zero and oscillates with the period of the drive. However, there is an additional  $\mathbb{Z}_2$ symmetry $a_j \rightarrow -a_j$, which is spontaneously broken when $\langle a_j \rangle \neq 0$, which occurs when the mode is Bose condensed. This leads to the appearance of the time-dependence $\langle a_j \rangle \propto e^{-\pi i \fD t}$ in the laboratory frame. The breaking of this $\mathbb{Z}_2$ symmetry thus leads to an oscillation with period $2\,\TD$. We propose that for low occupation ($\langle a_j \rangle \simeq 0$) the system is in a state dominated by a description based on the evolution of the pair correlation $\langle a_j a_j \rangle$. As occupation in the mode grows, \emph{i.e.}, the occupation number of the mode $\langle a_j \rangle$ goes up, there is a phase transition from the paired state to a state dominated by dynamics in $\langle a_j \rangle$, breaking the $\mathbb{Z}_2$ symmetry. We identify this transition as the phase transition to the time crystal.

In summary, we have shown the existence of a space-time crystal which is robust against fluctuations in experimental parameters and long-lived. Future experiments are aimed at studying elementary excitations such as solitons and sound in the presence of a space-time crystal, as our system is an excellent testing ground for these excitations. Moreover, it can be explored whether this spatially ordered state has supersolid properties, as this would allow study of out-of-equilibrium supersolids~\cite{leonard,li}, combining the fields of time crystals and supersolids and exploring a currently unknown corner of physics.

We thank Alexander Groot and Pieter Bons for their contribution to the initial stages of this research. This work is supported by the China Scholarship Council (CSC), the Stichting voor Fundamenteel Onderzoek der Materie (FOM) and is part of the D-ITP consortium, a program of the Netherlands
Organization for Scientific Research (NWO) that is funded by the Dutch Ministry of Education, Culture and Science (OCW).

\bibliographystyle{apsrev}

\clearpage

\section{Supplementary material}

\subsection{Sample preparation and excitation} \label{subsec:m_prep_and_excit} A Bose-Einstein condensate (BEC) of $^{23}$Na atoms is prepared in a cylindrically symmetric magnetic trap (see Fig.~1a of the Letter) with harmonic trapping frequencies $(\omega_\rho,\omega_z) = 2\pi\times (52.7,1.43)\,\textrm{Hz}$. The particle number is $N \simeq 5\times10^7$ and a condensate fraction of $N_0/N \simeq 90\%$ is reached. After sample preparation, the atom cloud is perturbed by modulating the radial trap frequency with 3 consecutive pulses of length $50\,\textrm{ms}$ and modulation depth $0.125$. This perturbation induces mainly a radial breathing oscillation of the condensate, which acts as a drive for the high-order axial excitation.

\subsection{Determining the line density} \label{subsec:m_linedens} Phase contrast imaging is used to make  50 images of a single atom cloud over a period of 160 ms allowing the study of dynamics of a single BEC. By combining multiple measurement series together, the dynamics of the atom cloud can be studied over many seconds with large time resolution. The signal from phase contrast imaging is given by~\cite{meppelink_pci}
\begin{equation}
	S(x,z) = 2 - 2 \cos\left[\pi/3 + \frac{\alpha}{2\varepsilon_0} n_\mathrm{c}(x,z)\right],
	\label{eq:m_pci_sign}
\end{equation}
where $\varepsilon_0$ is the vacuum permittivity and $\alpha$ is the polarizability of the atoms. The column density $n_\mathrm{c}(x,z)$ is related to the particle density denoted by $n(x,y,z)$ as $n_\mathrm{c}(x,z) = \int\!\mathrm{d}y\, n(x,y,z)$, where the integral runs over the propagation direction of the probe beam. The line density is obtained by integrating out the second radial direction to obtain 
\begin{equation}
	n_\ell(z) = \int\!\mathrm{d}x\, n_\mathrm{c}(x,z).
	\label{eq:m_linedens}
\end{equation}
During sample creation a slight axial center-of-mass motion is introduced. This center-of-mass motion is uncoupled but has to be corrected for to prevent the space-time crystal lattice to appear tilted. As a measurement run typically takes $160\,\mathrm{ms}$ and the axial trap frequency is $\omega_z/2\pi = 1.43\,\textrm{Hz}$, the motion can approximated to be linear within a single run. Line density profiles of consecutive images are shifted such that the center of the atom cloud is lined up for all images within a single run.

\subsection{Determining amplitude of drive and crystal} \label{subsec:m_ampl} To determine the relative amplitude of the driving mode $\delta \AD/\bAD$, the radial size of the atom cloud $R_\rho(t)$ is determined for every frame by performing a least squares fit  directly on the data using a Thomas-Fermi distribution~\cite{pethick_smith}. By assuming an oscillation of the form $R_\rho(t) \propto \bAD + \delta \AD\,\cos(2 \pi \fD t)$ and calculating the standard deviation $\sigma_\rho$ and mean $\bar{R}_\rho$ for a time interval $[t,t+\Delta t]$, the expression $\sqrt{2} \sigma_\rho/\bar{R}_\rho = \langle \delta \AD/\bAD \rangle$ yields the relative amplitude over this time interval. Here we choose $\Delta t = 50\,\mathrm{ms}$.

To determine the crystalline fraction for the space-time crystal, for every measurement run the first 30 images are selected and the center of the atom cloud is used as in Fig.~3a,c of the Letter. Subsequently for every measurement run the crystalline fraction is determined through the formula
\begin{equation}
	f_\mathrm{X} = \sqrt{\frac{\sum_{(k,f)\in I} |\mathcal{F}(n_\ell)(k,f)|^2}{\sum_{\mathrm{all}\,(k,f)} |\mathcal{F}(n_\ell)(k,f)|^2}} .
	\label{eq:m_crys_frac}
\end{equation}
The set $I$ is chosen as a union of 4 subsets $[\pm \kD -\Delta k,\pm \kD +\Delta k]\times[\pm \fD/2 -\Delta f,\pm \fD/2 +\Delta f]$ centered on the 4 peaks $(\pm \kD,\pm \fD/2)$. Here, $\Delta k$ is 3 points in Fourier space and is chosen such that any contributions from off-center $k$ as a result of lattice spacing variations due to density inhomogeneities are included. In addition, $\Delta f$ is chosen to be one point in Fourier space to account for a slight tilt of the crystal lattice due to imperfect compensated axial center-of-mass motion. This corresponds to $\Delta f\approx 0.05 \fD$. The dashed line in Fig.~4 of the Letter corresponds to the first measurement run without wait time where no pattern is observed. This is an indicator for the background level due to shot noise.

\subsection{Periodicity} \label{subsec:m_periodicity} The periodicity of the crystal is determined by the local speed of sound $c$ and the driving frequency $\fD$. Since the pattern oscillates at frequency $\fD/2$, the associated wavenumber with the excitation is $\kD = \pi \fD / c$. The speed of sound depends on the central density and is given by $ c = \sqrt{\bar{n} \TtwoB/m}$, where $m$ represents the sodium atom mass and $\bar{n}$ is the cross-sectional averaged density in the center of the trap, given by $\bar{n} = n(0,0,0)/2$ where $n(0,0,0)$ is the particle density (see, for example Ref.~\cite{zaremba}). For the experimental data in Fig.~3b,d of the Letter, this yields $\kD = 0.110\,\mu\textrm{m}^{-1}$. For the simulation data in Fig.~3f,h, this yields $\kD = 0.096\,\mu\textrm{m}^{-1}$.



\subsection{Collective modes} We briefly discuss the most important collective modes present in our experiment and identify coupling between different modes. The cloud of $^{23}$Na is trapped in a cylindrically symmetric harmonic trap with trapping frequencies $(\wrho,\wz) = 2\pi\times (52.7,1.43)\,\textrm{Hz}$, or $\wrho \approx 40\,\wz$. The radial directions are referred to as $\hat{x}$ and $\hat{y}$, while the axial direction is referred to as $\hat{z}$. Imaging is performed along the $\hat{y}$-direction, which is the radial direction coinciding with that of gravity.

\subsubsection{Dipole motion} The dipole motion is the center-of-mass motion of the atomic cloud. In a harmonic trap  the dipole motion is undamped as dictated by Kohn's theorem~\cite{Kohn}, does not couple to any other mode and does not influence the dynamics of the cloud. From the dipole motion the trapping frequencies in the imaging plane can be determined with large accuracy and the frequency along the imaging direction in the cylindrically symmetric trap is identical to the radial direction in the image plane.

\subsubsection{Radial breathing mode} The radial breathing mode~\cite{collective_oscillations_stringari,monopole_dalibard}, also referred to as the radial quadrupole mode, is an oscillation of the condensate width in the radial direction with a small out-of-phase oscillation of the length of the condensate in the axial direction. The mode acts as the drive for the time crystal and has a frequency of approximately $ 2\,\wrho$. This mode is strongly excited with relative amplitude $\langle \delta \AD/\bAD \rangle \approx 0.3$ and is thus in the non-linear regime.

\subsubsection{Axial breathing mode} The axial breathing mode, also referred to as the axial quadrupole mode, is an oscillation of the condensate length in the axial direction complemented by an out-of-phase oscillation of the radial width. During the decompression of the trap after the cooling process this mode is excited in the experiment. The frequency associated with this mode is about $1.55\,\wz$  and is low compared to all the other frequencies in the experiment. The effect of the mode on the experimental results is small and is taken into account as adiabatic changes of the condensate size.

\subsubsection{Scissors mode} The scissors mode~\cite{scissor_foot} is a tilting of the condensate in the trap and its frequency is given by $\sqrt{\wz{}^2+\wrho{}^2}$. The mode is weakly coupled to the dipole mode due to terms in the potential of the form $V(x,y,z) \propto xz$ resulting from imperfections in the magnetic field used for trapping the atoms. It also couples to the radial breathing mode,  which is related to the collapse and revival of the radial breathing mode. A detailed analysis of this coupling is given in the section below.

\subsubsection{Higher-order axial modes} The collection of higher-order axial modes are of crucial importance for this Letter. Their frequencies are approximately given by $\wz \sqrt{(4j+1)(4j+2)} / 2$ and their mode function is given by $P_{4j+2}(\tilde{z}) - P_{4j}(\tilde{z})$ with $\tilde{z} = z / R_z(t)$ and $P_j$ the $j$'th Legendre polynomial.  They are strongly coupled to the radial breathing mode.

\subsection{Scissor mode in the images}
As mentioned in the caption of Fig. 3 of the Letter, the peaks at $k=0$ and $f/\fD = \pm 1$ are associated with the scissor mode. The angle of the cloud with the equilibrium causes the atom cloud to  appear shorter but denser in the density profiles. This effect is independent of the sign of the angle and leads to peaks in Fourier space at double the frequency of the scissor mode. So for a scissor mode at $f_\mathrm{sc} = \fD/2$, this will appear in the line density spectrum at $f/\fD = \pm 1$, as in Fig. 3b,d. As there is no scissor mode present in the simulation, these peaks are absent in Fig. 3f,h.

\subsection{Numerical simulations}
Simulations  are performed using the time-splitting spectral method, of which a good description is given by Bao \emph{et al.}~\cite{bao}. For the simulations presented here the cylindrical symmetry is exploited by writing the condensate wavefunction as $\psi(\rho,z,\varphi,t) = \Phi(\varphi) f(\rho,z,t)/\sqrt{\rho}$ and  $\Phi(\varphi) = 1/\sqrt{2\pi}$ is assumed to be constant. The equation of motion for $f(\rho,z,t)$ derived from the Gross-Pitaevskii equation is given by
\begin{widetext}
\begin{equation}
	(i-\gamma)\hbar \partial_t f(\rho,z,t) = \left( -\frac{\hbar^2}{2m} \left[\partial_\rho^2  + \partial_z^2 -\frac{1}{4 \rho^2} \right]+V(\rho,z)+\frac{\TtwoB}{2 \pi \rho} |f(\rho,z,t)|^2 - \mu \right) f(\rho,z,t),
	\label{eq:suppl_sim_1}
\end{equation}
\end{widetext}
where the non-linear interaction parameter $\TtwoB$ is related to the $s$-wave scattering length $a$ through $\TtwoB=4\pi a \hbar^2/m$, $\mu$ is the chemical potential and $m$ is the atomic mass. This is a two-dimensional Gross-Pitaevskii equation with a contact interaction strength dependent on $\rho$. The damping constant $\gamma$ is chosen to be $7 \times 10^{-4}$, which damps out strong gradients as a result of numerical errors, but does not affect the dynamics. The harmonic potential is  given by 
\begin{equation}
V(\rho,z) = \frac{1}{2} m \left(\wrho{}^2 r^2 + \wz{}^2 z^2\right). \label{eq:trap}
\end{equation} 
The simulations are performed on a $256 \times 1024$ point grid with total physical grid size $[ -2\,R_\rho,2\,R_\rho] \times [ -2\,R_z,2\,R_z]$, where $R_i = \sqrt{2\,\mu / (m \omega_i{}^2)}$ is the equilibrium Thomas-Fermi radius in the direction $i$ = $\rho$, $z$. 

At the start of the simulation the wavefunction is initiated in the Thomas-Fermi profile. Subsequently, imaginary-time evolution is performed on $f(\rho,z,t)$ to decay to the ``true'' ground state. The imaginary-time evolution is continued until the energy of the system is constant within numerical accuracy. For the study of the excitations in the system,  real-time evolution is performed on $f(\rho,z,t)$ by applying the same sequence on the trap frequencies as in the experiment. By modulating $\wrho$ with a certain modulation depth, the amplitude of the radial breathing mode can be tuned. This allows for the study of the dynamics of the modes for several hundred radial trap periods, which can be compared to the experimental results.


\begin{figure}[!t]
	\centering
	\includegraphics[width=85mm]{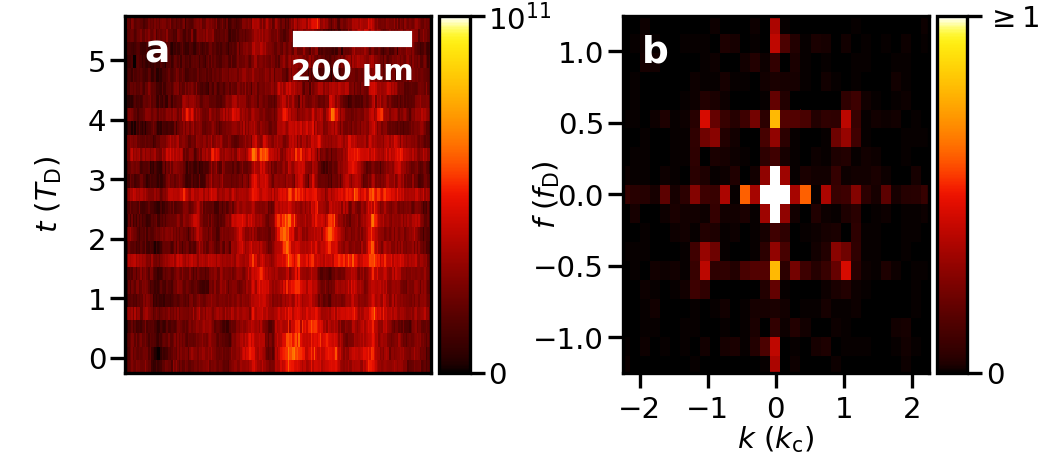} 
	\caption{(Color online) \textbf{(a)} Line density $n_\ell$ as a function of time and position starting $200\,\mathrm{ms}$ after the onset of the drive. The trap frequencies are $(\omega_\rho,\omega_z) = 2\pi\times (28.4,1.39)\,\textrm{Hz}$ and the number of atoms $N \simeq 1.6\times10^7$.  \textbf{(b)}   Fourier transform of the data in (a), where the peaks at $(k/\kD,f/\fD) = (\pm 1,\pm 0.5)$ are associated with the excitation pattern. Here $\kD$ is determined using the speed of sound $c$ for the experimental parameters used in the experiment. \label{fg:six}}
\end{figure}

\subsection{Frequency of the pattern}

We have carried out measurements, where the axial excitation pattern is excited under different experimental conditions by using different radial and axial trap frequencies, short and long kicks of the system, different temperature of the cloud and different number of atoms. Although the pattern appeared at different times after the initial kick, it always showed up. As an example, Fig.~\ref{fg:six} shows the result for $(\omega_\rho,\omega_z) = 2\pi\times (28.4,1.39)\,\textrm{Hz}$ and $N \simeq 1.6\times10^7$. Since the number of images analyzed is fewer compared to Fig.~2 of the Letter, the resulting Fourier peaks are broader. However, the frequency is again $f_{\rm D}/2$. 


\subsection{The coupling between the radial breathing mode and the scissors mode\label{sc:coup}}

Collective modes in an atomic Bose-Einstein condensate are described by the time-dependent Gross-Pitaevskii equation. We consider a radial symmetric harmonic trap given by Eq. (2), where the trap anisotropy is given by $\lambda\equiv\wz/\wrho\ll 1$. To study the $xz$-scissors mode and its coupling to the radial breathing mode in the trap, we use the following Gaussian trial function for the
condensate wavefunction~\cite{Khawaja,Henk}
\begin{eqnarray}\label{eq2}
\psi(\vr,t) & = & \sqrt{N} A(t) \times \\ 
& & \exp\left(-b_\rho(t)(x^2+y^2)-b_z(t)z^2 -c_{xz}(t)xz\right), \nonumber
\end{eqnarray}
where $b_\rho$, $b_z$ and $c_{xz}$ are time-dependent, complex  variational parameters with $b_\rho=b_{\rho,r}+ib_{\rho,i}$, $b_z=b_{z,r}+ib_{z,i}$ and $c_{xz}=c_{xz,r}+ic_{xz,i}$. Furthermore, the amplitude $A(t)$ of the wavefunction is defined as 
\begin{equation}\label{eq3}
A(t)= \frac{1}{\pi^{3/4}} \left( 2b_{\rho,r}(4b_{\rho,r}b_{z,r}-c_{xz,r}{}^2) \right)^{1/4},
\end{equation}
which guarantees that the wavefunction remains at all times normalized to the total particle number $N$. The parameters $b_\rho$ and $b_z$ determine the radial breathing mode with $b_{z,r}\ll b_{\rho,r}$. The parameter $c_{xz}$ describes the scissors mode in the $x$-$z$ plane. To derive the equations of motion for these variational parameters, we consider the Lagrangian of the Gross-Pitaevskii equation,
\begin{widetext}
\begin{equation}\label{eq4}
L[\psi,\psi^*]  =  \frac{1}{2}i\hbar\int d\vr\left(\psi^*\frac{\partial \psi}{\partial t}-\psi\frac{\partial \psi^*}{\partial t}\right) 
 - \int d\vr\left(\frac{\hbar^2}{2m}|\nabla\psi|^2+(V(\vr)-\mu)|\psi|^2+\frac{1}{2}\TtwoB|\psi|^4 \right).
\end{equation}
\end{widetext}
We introduce the oscillator length $\ell=\sqrt{\hbar/m\bar{\omega}}$ associated with the geometrical average frequency $\bar{\omega}=(\wrho{}^2 \wz)^{1/3}$.  Substituting our trial wavefunction from Eq.~(\ref{eq2}) into the Lagrangian of Eq.~(\ref{eq4}), we obtain
\begin{widetext}
\begin{eqnarray}\label{eq5}
\frac{L[b_\rho,b_z,c_{xz}]}{N} &=& \frac{1}{2\pi^2 A^4} \left\{ \left[ \rule[-4mm]{0mm}{8mm} 4b_{\rho,r}{}^2\dot{b}_{z,i}+8b_{\rho,r}b_{z,r}\dot{b}_{\rho,i}-2b_{\rho,r}c_{xz,r}\dot{c}_{xz,i}\right] \right. \nonumber\\
&&-{2} \left[ \rule[-4mm]{0mm}{8mm}  4b_{\rho,r}\left( 2b_{z,r}b_{\rho,i}{}^2+b_{\rho,r}\left(b_{z,i}{}^2+b_{z,r}(2b_{\rho,r}+b_{z,r})\right)\right) +b_{\rho,r}(b_{\rho,r}+b_{z,r})c_{xz,i}{}^2 \right.\nonumber\\
&&\left. \rule[-4mm]{0mm}{8mm}  -2b_{\rho,r}(b_{\rho,i}+b_{z,i})c_{xz,r}c_{xz,i} -\left(b_{\rho,i}{}^2+b_{\rho,r}(2b_{\rho,r}+b_{z,r})\right)c_{xz,r}{}^2\right]
\nonumber\\
&&-\left. \left[ \rule[-4mm]{0mm}{8mm} 4b_{\rho,r}b_{z,r}\wrho{}^2+2b_{\rho,r}{}^2\wz{}^2-\frac{1}{2}c_{xz,r}{}^2\wrho{}^2\right]\right\}
- \gamma \sqrt{\frac{\pi}{2}} A^2 ,
\end{eqnarray}
\end{widetext}
where we introduce dimensionless variables by scaling all lengths with $\ell$ and all times with $1/\bar{\omega}$. Here $\gamma=Na/\ell$ is the dimensionless parameter that determines the strength of the interaction. The first term within square brackets in Eq.~(\ref{eq5}) derives from the time-derivative of Eq.~(\ref{eq4}), the second from the kinetic energy, the third from the potential energy, and the last term from the interaction term. Using the Euler-Lagrange equations for all variational parameters, we derive the coupled equations of motion. To compare with experiment we are especially interested in the variables $q_{\rho}=1/\sqrt{2b_{\rho,r}}$ and $q_{z}=1/\sqrt{2b_{z,r}}$ that correspond to the radial and axial width  of the condensate. The equilibrium values $\bar{q}_\rho$ and $\bar{q}_z$ are determined by~\cite{pethick_smith} $\bar{q}_\rho=(2\gamma^2/\pi)^{1/10}/\wrho$ and $\bar{q}_z=(2\gamma^2/\pi)^{1/10}/\wz$.  The rotational angle $\theta$ of the scissors mode in the $x$-$z$ plane is determined by the relation $\theta=q_{\rho}{}^2q_{z}{}^2c_{xz,r}/(q_{z}{}^2-q_{\rho}{}^2)$.

\begin{figure}
	\centering
	\includegraphics[width=89mm]{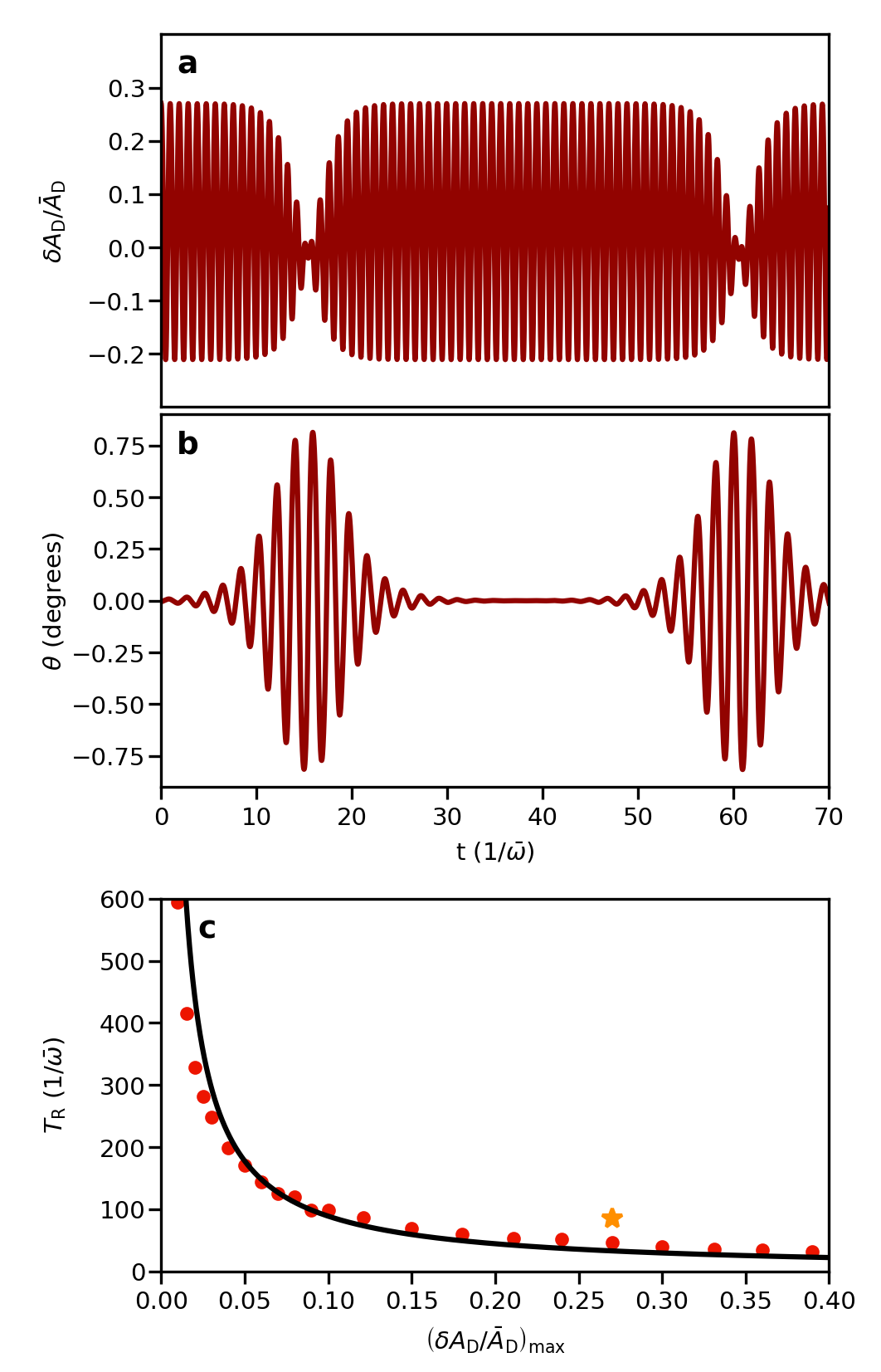} 
	\caption{(Color online) Results of solving the coupled equations of motion numerically. (\textbf{a}) The radial width of the condensate, which shows the time-dependence of the radial breathing mode. (\textbf{b}) The angle $\theta$ of the condensate, which shows the time-dependence of the scissors mode. (\textbf{c}) The revival period $\TR$ from the simulation (dots), the approximated result (solid line) using the function $\TR=\eta/[\wrho (\delta \AD/\bAD)_{\rm max}]$ with $\eta=30$, and the result from the experiment (star). \label{figsix}}
\end{figure}

The results of solving the equations of motion numerically for the experimental parameters of Fig.~4 of the Letter are shown in Fig.~\ref{figsix}. We find that the coupling between the $xz$-scissors mode and the radial breathing mode induces a collapse and revival of the radial breathing mode. We observe these collapses and revivals only when the radial breathing mode is in the nonlinear regime, since in the linear regime the $xz$-scissors mode and the radial breathing mode are uncoupled. Applying the rotational wave approximation we can simplify the equations of motion to find that the revival period $\TR$ is determined by $\TR=\eta/[\wrho (\delta \AD/\bAD)_{\rm max}]$,  where $(\delta \AD/\bAD)_{\rm max}$ is the maximum of $(\delta \AD/\bAD)=q_{\rho}(t)/\bar{q}_\rho-1$ and $\eta$ is an adjustable parameter determined from  the numerical solutions. Although there is no damping in our model and given the variational nature of our approach, the prediction for $\TR$ shown in Fig.~\ref{figsix} is reasonably close to the experimental result.

\end{document}